# Influence of anisotropic magnetoresistance on nonlocal signals in Si-based multi-terminal devices with Fe electrodes

Ryosho Nakane, Shoichi Sato, Shun Kokutani, and Masaaki Tanaka

Department of Electrical Engineering and Information Systems, The University of Tokyo, 7-3-1 Hongo Bunkyo-ku, 113-8656 Tokyo, Japan

Abstract

We have investigated the influence of anisotropic magnetoresistance (AMR) on nonlocal signals in Si-based multi-terminal devices with ferromagnetic Fe electrodes. The AMR of the Fe electrodes was found to have a significant influence on nonlocal signals when the in-plane device structure is not optimized. Moreover, realization of a pure spin current by spin diffusion was found to be virtually impossible because of the electric potential distribution in the depth direction in the Si channel. Although apparent signals indicating the spin-valve effect were not detected, we mainly present structural influence on the electric potential distribution which is indispensable for the analyses of spin-dependent transport.

Spin-functional semiconductor devices are expected to be good candidates for key devices in next-generation integrated circuits.[1-5] For realizing such devices, it is necessary to clarify the physics of spin-dependent transport in semiconductors.[6-15] Electrical measurements in the non-local geometry[8-15] (hereafter referred to as the nonlocal measurement) are commonly employed, because, if ideally realized, a pure spin current induced by spin diffusion enables us to estimate the important physical parameters, such as spin diffusion length, with excluding unwanted charge-current-related phenomena including the anisotropic magnetoresistance (AMR)[16] and local Hall effect.[17,18] The purpose of this study is to experimentally evaluate the pure spin current in the nonlocal measurement with focusing on the difference in the device structure. While only the magnitude of the nonlocal voltage change has been discussed so far, there is no report, to our knowledge, on the detailed analysis of the nonlocal measurement in multi-terminal spin devices with a semiconductor channel.

We prepared substrates with a thermally-oxidized 30-nm-thick $SiO_2$ surface layer and a 90-nm-thick Si channel layer, which were fabricated using a silicon-on-insulator wafer with a 200-nm-thick buried oxide (BOX) layer. Figure 1(a) shows the schematic design of our device structure examined in this study, in which the $x$, $y$, and $z$ coordinates are defined. Hereafter, the edge-rounded 20-nm-thick ferromagnetic Fe electrodes with areas of $2 \times 180$ μm$^2$ (the left side) and $10 \times 180$ μm$^2$ (the right side) are referred to as FM1 and FM2, respectively, and the outside two nonmagnetic contacts are denoted as NM1 and NM2. The important device parameters are as follows: The thickness of a Ta capping on Fe is 3 nm, the phosphorous doping density in the n-Si channel is $2 \times 10^{19}$ cm$^{-3}$, the gap length $L_{GAP}$ between the two Fe electrodes is varied from 3 to 20 μm, the distance $L_{REF}$ between the FM1 (FM2) and NM1 (NM2) is 100 μm, and the channel widths along the $y$ direction are 154 μm for type-I and 184 μm for type-II, respectively. All the metal electrodes and pads were fabricated using electron-beam (EB) lithography, EB evaporation, thermal evaporation, and lift-off. In the fabrication process, the surface $SiO_2$ layer on Si in the contact areas were removed with buffered HF just before the electrode deposition. Finally, the island device structure on the BOX layer was formed using reactive ion etching with $CF_4$ and $SF_6$ gases. We examined two types of devices (type-I and II) as illustrated in Fig. 1(b) and (c), respectively, where each pad is named as A – H. All the devices have FM1, FM2, NM1, and NM2, while the structure and number of the Al pads attached to FM1 and FM2 are different. The Fe electrode (FM1, FM2) in type-I has a pad (B or C) at the center and two pads (E, G or F, H) connected to the Fe electrode to measure the AMR signals of FM1 and FM2 with a two probe method. Note that the three pads for each Fe (named by B, E, G, and C, F, H in Fig. 2(b)) are electrically connected only through Fe, and thus the surface of the Fe electrodes is not fully covered. Type-II has a simple structure, in which square B and C pads fully cover the Fe electrodes.

We mainly describe our experimental results at room temperature for devices with $L_{GAP} = 3$ μm otherwise noted, and the magnetoresistance was measured with sweeping an in-plane magnetic field applied along the $y$ direction which is the easy magnetization axis of the Fe electrodes. *I-V* characteristics through the Si channel were almost linear due to the highly-doped Si channel. The estimated channel resistances for $L_{REF} = 100$ μm are 333 Ω for type-I and 400 Ω for type-II, respectively. The estimated Schottoky junction resistances for FM1 ($R_{J1}$) and for FM2 ($R_{J2}$) are 80 Ω and 50 Ω for type-I and 320 Ω and 99 Ω for type-II, respectively. Figure 2(a) shows the AMR of FM1 and FM2 measured at room temperature with a constant current of 100 μA driven between E and G (F, and H). The abrupt voltage changes are attributed to the magnetization reversal of FM1 and FM2, whose coercivities are 100 ± 2.5 Oe and 40 ± 2.5 Oe, respectively.

Our typical procedure for the nonlocal measurement is to detect the voltage between C (grounded) and A, B, E, G while a constant current flows between C and D, which is shown in Fig. 1(c). This procedure makes it easy to obtain the voltage and current distributions, and to estimate the nonlocal signals by subtraction. Hereafter, the voltage of each pad is referred to as $V_\alpha$ ($\alpha$ = A–H), the nonlocal voltage ($V_B - V_A$) is referred to as $V_{NON}$, and the voltage change in magnetoresistance is represented by Δ.  First, the nonlocal measurement was performed for type-I devices.  Figure 2(b) and (c) show $V_A$ and $V_B$ with a current of 1 mA, respectively, while electrons were injected from C to D.  In these figures, both $V_A$ and $V_B$ reflect the AMR of both FM1 and FM2, indicating that some amount of electrons goes through FM1 and FM2 along the $y$ direction. It is noteworthy that the signal originating from FM1 is negative for $V_A$ whereas that is positive for $V_B$ (dotted circles in Fig. 2(b) and (c)), but the reason for the opposite polarity is not clear.  Since $V_A$ and $V_B$ are different, $\Delta V_{NON}$ estimated by subtracting Fig. 2(b) from Fig. 2(c) exhibits hysteretic behavior, as shown in Fig. 2(d), where $V_{OFFSET}$ (the minimum voltage) is -4.721 mV.  Since the positive change in Fig. 2(d) between 40 and 100 (-40 and -100) means the decrease in the resistance Oe, the polarity of the voltage change contradicts the spin-valve effect.  Note that the polarity of the signals in Fig. 2(b) – (d) was inversed when the current direction was inversed.

To confirm whether the spin-valve effect due to spin injection and detection appears or not, a minor loop was measured; first, the magnetic field was swept from 200 Oe to -60 Oe (the valley in Fig. 2(d)) and from -60 Oe to 200 Oe.  Since the minor loop of $\Delta V_{NON}$ (dotted curve in Fig. 2(d)) reflected only the AMR of FM2, the spin-valve effect is unlikely to have appeared.  It was also found that $\Delta V_{NON}$ linearly increased with increasing the current in the range of -4 to 4 mA (not shown here).  On the other hand, when a measurement with the same procedure was performed for type-I devices without FM1 but with a nonmagnetic Al contact instead, nearly the same signal as Fig. 2(b) with a different magnitude was obtained for $V_A$ and $V_B$, leading to the hysteretic $\Delta V_{NON}$ which reflects the AMR of FM2.  Furthermore, in the devices with and without FM1, $\Delta V_{NON}$ was greatly enhanced when electrons were injected from H to D.  Since the spin-valve signal would not be changed by changing the pad for electron injection, the observed $\Delta V_{NON}$ originates from the AMR due to the electron flow in FM1 and FM2 along the $y$ direction.  Note that almost identical $\Delta V_{NON}$ with a slightly smaller magnitude was also obtained in the temperature range of 10 − 250 K.  On the other hand, in the nonlocal measurement with sweeping a perpendicular magnetic field, we did not observe apparent Hanle oscillation or Lorentz-magnetoresistance-like change as often seen in other reports[9-12,14] at any temperatures. From these results, we conclude that $\Delta V_{NON}$ in Fig. 2(d) originates from the AMR of FM1 and FM2. It is noteworthy that when the maximum change in $\Delta V_{NON}$ at ±40 Oe and $V_{OFFSET}$ were plotted as a function of $L_{GAP}$ (= 3, 5, 10, 20 μm), they showed exponential dependence, seemingly leading to the spin diffusion length of 9.1 μm using Eq. (8) in ref. 20, which is very similar to the reported values.[12,13] Thus, this exponential decrease in the in-plane $\Delta V_{NON}$ with $L_{GAP}$ is not enough for the evidence of the spin-valve effect.

On the other hand, for type-II devices in the nonlocal measurement with a constant current of 1 mA driven between C and D, $V_A$ and $V_B$ were almost the same, ~100 mV, and $V_{OFFSET}$ was ~ -2 μV. Since the voltage of the FM2/Si Schottky junction is estimated to be 99 mV = 1 mA × $R_{J2}$ (= 99 Ω), $V_A$ and $V_B$ are nearly equal to the surface voltage of the Si channel just below FM2. Whereas this leads to the complete suppression of AMR, $\Delta V_{NON}$ was undetectable (below 10 nV). Thus, if the spin-valve signal appears in those devices, its magnitude is below 10 nV. We anticipate that the formation of disordered alloys due to the reaction of Fe with Si affects the spin injection/detection efficiency.[21]

Using the voltages for all the pads, we quantitatively estimated the surface potential distribution on the channel of the type-I device when a constant current of 1 mA was driven between C and D, as shown in Fig. 3(a). Ideally, all the electric potential lines in the local region are parallel to the $y$ direction, whereas no electric potential line is present in the nonlocal region. Interestingly, however, the difference between $V_C$ and $V_F$ ($V_C$ and $V_H$) directly indicates the electron flow along the $y$ direction in FM2 from C to F (C to H). Also, the difference between $V_B$ and $V_E$ ($V_B$ and $V_G$) indicates the electron flow along the $y$ direction in FM1 from E to B (G to B). Since the AMR of FM2 (FM1) can change the electron flow in FM2 (FM1), it leads to the change in the potential distribution in the channel. The electric field concentration at the edges of pad C (white dotted circles in Fig. 3(a)) is probably an origin of the electric field in the nonlocal region and the AMR influence on $V_{NON}$. Note that the electric potential lines at around pad C in Fig. 3(a) was drawn based on the calculated result for a two-dimensional edge structure using the conformal mapping method.[22,23]

To obtain a valuable insight into the electrical states in the Si channel, we further calculated two-dimensional lines of electric force and electric potential distributions, as shown in Fig. 3 (b), with the conformal mapping method.[22,23]  We investigated electrical states in the *x-z* plane of a Si channel at around a source FM electrode (FM2) by changing the aspect ratio *W*/*D* in the complex *x-iz* plane, where *D* and *W* are the Si channel depth and the FM electrode width, respectively. The structure is defined by $-\infty \leq x \leq \infty$ and $-iD \leq z \leq 0$, and the current ($\propto$ electric field) between the FM electrode and the Si channel was fixed at a constant value at $x \geq D$ for all the cases. Figure 3(b) shows the calculated electrical state for *W*/*D* = 0.1, where the dotted and solid lines represent electric potential lines and lines of electric force lines, respectively. To analyze the result further, a voltage distribution *V*(*x*) at the *z* boundaries is plotted as a function of position *x*, as shown in Fig. 3(c), where the solid, broken, dotted lines represent $V_0(x)$ from (0, 0) to (*D*, 0), $V_1(x)$ from $(-\infty, 0)$ to (-*W*, 0), and $V_2(x)$ from $(-\infty, -iD)$ to (*D*, -*iD*), respectively. When a detecting FM electrode locates near *x* = -*W*, $V_{OFFSET} = V_1(-(W + L_{GAP})) - V_1(-\infty)$ exhibits the exponential-like behavior with the change of $L_{GAP}$ and its polarity is negative. In Fig. 3 (d), we plot $V_1(-\infty)/V_0(D)$, which indicates the nonlocal voltage (and the drift current in the nonlocal region) normalized by the local voltage, as a function of *W*/*D*. It was found that $V_1(-\infty)/V_0(D)$ decreases exponentially with increasing *W*/*D*, and thus large *W*/*D* (> ~1) is favorable to obtain a pure spin current without charge current. Next, we evaluated the detection of the spin-dependent nonlocal voltage $\Delta V_{SPIN}$ (= $I \times \Delta R$, where *I* is a constant current and $\Delta R$ is defined by Eq. (23) in ref. 24) by another FM electrode at $x = -(W + L_{GAP})$. Since the spin polarization of electrons decreases exponentially with increasing distance due to the spin relaxation in a semiconductor channel, large $\Delta V_{SPIN}$ is expected when a large amount of injected spin-polarized electrons diffuses in a short distance below the spin diffusion length.

The electron current $i(x)$ injected from the FM at position ($-W < x < 0$, $z = 0$) was calculated, assuming that it is proportional to the electric field just below the FM. The distribution of $i(x)$ shows a monotonic decrease from $x = 0$ to $-W$ when $W/D \geq \sim 1$, whereas it bends downward with a minimum at $x = \sim -0.5W$ when $W/D \leq \sim 0.1$. Then, we integrated $i(x)$ to obtain a partial current $I$ with each range; $I_W$ ($-W < x \leq -0.9W$), $I_M$ ($-0.6W \leq x < -0.5W$), $I_0$ ($-0.1W \leq x < 0$), and the total current $I_{TL}$ ($-W < x < 0$). The calculated values are as follows: $I_W = 0.02 I_{TL}$, $I_M = 0.03 I_{TL}$, and $I_0 = 0.50 I_{TL}$ for $W/D = 2$, $I_W = 0.18 I_{TL}$, $I_M = 0.06 I_{TL}$, and $I_0 = 0.26 I_{TL}$ for $W/D = 0.1$. Owing to the electric field concentration at the edges ($x = -W, 0$), $I_0$ is largest in any $W/D$ and $I_W$ strongly depends on $W/D$. As a result, although large $W/D$ ($\geq \sim 1$) is favorable for the pure spin current, this leads to small $\Delta V_{SPIN}$ since $I_0$ dominates $I_{TL}$ and the electrons in $I_0$ should travel a relatively long distance $W + L_{GAP}$. On the other hand, large $\Delta V_{SPIN}$ is expected for small $W/D$ ($\leq \sim 0.1$) since $I_W$ is comparable to $I_0$ and the electrons in $I_W$ should travel a relatively short distance $L_{GAP}$. However, for small $W/D$ the electron transport in the nonlocal region is no longer a pure spin current because it contains a significant drift current contribution, as shown by the lines of electric force in the nonlocal region in Fig. 3(b).

To evaluate the magnitude of the electric field in the nonlocal region, one feasible method is to compare the voltage drop of the source/channel junction $V_J$ with $V_B$ and $V_A$ in Fig. 1(c). If $V_B = V_J \pm 0.5 \times \Delta V_{SPIN}$ in antiparallel/parallel magnetization configurations (the spin-valve effect) and $V_{OFFSET} = V_J - V_A = 0$, there is no electric field and a pure spin current is realized. In our experiments, $\Delta V_{SPIN}$ was not obtained, thus the required condition changes to $V_J = V_B$ and $V_{OFFSET} = 0$. Using this method, type-II was found to reasonably fulfill the requirements.

In summary, there is significant influence of the AMR of the FM electrodes on the nonlocal signals in our type-I devices. Such influence is excluded in our type-II devices, but the electric field in the nonlocal region is still present. This is explained by the electric potential distribution in the depth direction in the Si channel (Fig. 3(b)). Thus, the voltages in the nonlocal region should be measured and taken into account when the nonlocal signals are analyzed. Otherwise, one cannot precisely estimate the parameters in spin-dependent transport using the commonly used spin diffusion theory.

Acknowledgements

This work was partly supported by Grant-in-Aids for Scientific Research (including Young Scientists (A), challenging Exploratory Research, and Specially Promoted Research), the Special Coordination Programs for Promoting Science and Technology, the FIRST Program of JSPS.

Figure captions

Figure 1 (a) Schematic structure of our device, where the edge-rounded Fe electrodes with 2 × 180 μm$^2$ in area (the left side) and 10 × 180 μm$^2$ in area (the right side) are referred to as FM1 and FM2, respectively, and the outside two nonmagnetic contacts are denoted as NM1 and NM2. The *x*, *y*, and *z* coordinates are defined in the figure. (b)(c) Top view of (b) type-I and (c) type-II device structures, where each pad is named as A − H. (b) E, B, and G pads (F, C, and H pads) are electrically connected only through FM1 (FM2). (c) Typical geometry for the nonlocal measurement, in which a charge current is driven along the closed arrow in the local region. Theoretically, a pure spin current due to spin diffusion is induced along the open arrow in the nonlocal region, which can be detected as the voltage difference between $V_A$ and $V_B$.

Figure 2 (a) Anisotropic magnetoresistance (AMR) of FM1 and FM2 in a type-I device measured at room temperature, where a constant current of 100 μA was driven between E and G for FM1 (F and H for FM2). A sweeping in-plane magnetic field was applied along the *y* direction. (b)(c) Voltage changes $\Delta V_A$ and $\Delta V_B$ vs. magnetic field measured at room temperature with a current of 1 mA, where electrons are injected from C to D. The broken and solid lines are signals for different sweep directions (open and solid arrows) of the in-plane magnetic field along the *y* direction. The offset voltages are also shown in the figures, and the signal originating from FM1 is indicated by dotted circles. (d) Nonlocal signal $V_{NON}$ and its change $\Delta V_{NON}$ estimated by subtracting (b) from (c), where the nonlocal offset voltage $V_{OFFSET}$ is -4.721 mV. The broken and solid lines are signals for the different sweep directions as in (b) and (c). The minor loop is also shown by the thin dotted curves in the figure.

Figure 3 (a) Schematic in-plane electric potential lines on the surface of the Si channel in the type-I device, where the two dotted lines (the two broken lines) are the same potential. The estimated voltages of all the pads are shown in mV unit in the figure when a current of 1 mA was driven between C and D. The red and pale red regions represent the FM electrodes (FM1, FM2) and the middle pads (C and D), respectively. The white dotted circles denote the electric field concentration at the edges of pad C. (b) Distribution of equipotential lines (dotted lines) and lines of electric force (solid lines) in the complex $x$-$iz$ plane calculated by the conformal mapping method, where the device structure is defined by $-iD \leq z \leq 0$. The figure shows the result for $W/D = 0.1$, where $D$ and $W$ are the Si channel depth and the width of the FM electrode, respectively. The source FM electrode (the red square) was drawn after the calculation. (c) Surface potential distributions $V_0(x)$, $V_1(x)$, and $V_2(x)$ for $W/D = 0.1$ as a function of position $x$, where the solid, broken, dotted lines represent $V_0(x)$ from $(0, 0)$ to $(D, 0)$, $V_1(x)$ from $(-\infty, 0)$ to $(-W, 0)$, and $V_2(x)$ from $(-\infty, -iD)$ to $(D, -iD)$, respectively. (d) $V_1(-\infty)/V_0(D)$ as a function of $W/D$, in which the electric field at $x \geq D$ was fixed at a constant value for all the cases.

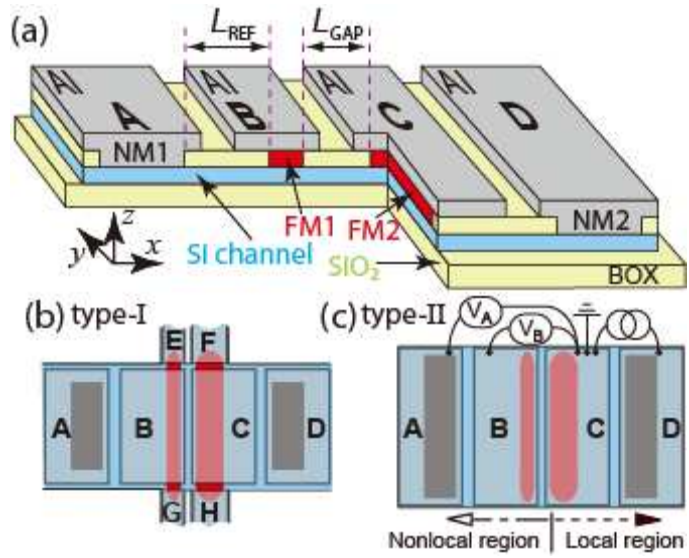

Fig.1　Nakane et al.

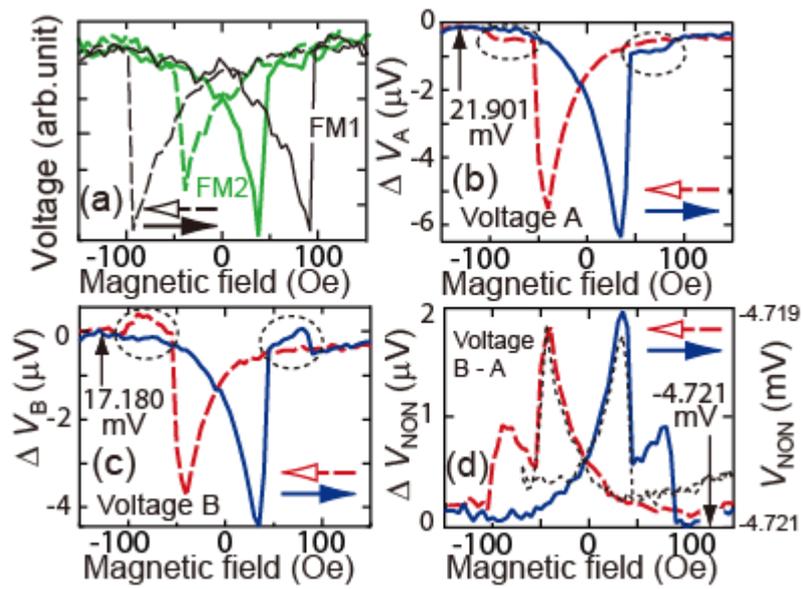

Fig.2　Nakane et al.

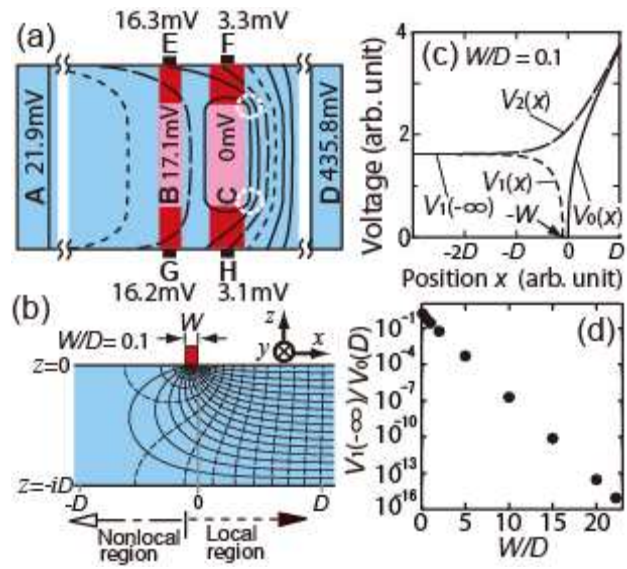

Fig.3   Nakane et al.